\documentclass{article}

\usepackage {graphicx,latexsym}
\usepackage{amsmath,amsthm}
\usepackage{amsfonts}
\usepackage{amssymb}
\usepackage[utf8]{inputenc}
\usepackage{authblk}
\usepackage{mathtools}

\usepackage{caption}
\usepackage{subcaption}

\begin{document}






\title{A better space of generalized connections}
\author[1]{
Juan Orendain
\footnote{e-mail: \ttfamily juan.orendain@case.edu}
}
\author[2]{
José A. Zapata
\footnote{e-mail: \ttfamily zapata@matmor.unam.mx}
}
\affil[1]{
Case Western Reserve University, USA.
}
\affil[2]{
Centro de Ciencias Matemáticas, 
Universidad Nacional Autónoma de México, 
C.P. 58089, Morelia, Michoacán, México.
}

\date{}
\maketitle

\begin{abstract} 
Given a base manifold $M$ and a Lie group $G$, 
we define $\bar{\cal A}^H_M$ a space of generalized $G$-connections on $M$ with the following properties: 

- The space of smooth connections ${\cal A}^\infty_M = \sqcup_\pi {\cal A}^\infty_\pi$ is densely embedded in $\bar{\cal A}^H_M = \sqcup_\pi \bar{\cal A}^H_{cc(\pi)}$; moreover, in contrast with the usual space of generalized connections, the embedding preserves topological sectors. 

- $\bar{\cal A}^H_M$ is a homogeneous covering space for the standard space of generalized connections of loop quantization $\bar{\cal A}_M$. 

- $\bar{\cal A}^H_M$ is a measurable space that can be constructed as an inverse limit of of spaces of connections with a cutoff, much like $\bar{\cal A}_M$. At each level of the cutoff, a Haar measure, a ``BF measure'' and heat kernel measures can be defined. 

- The topological charge of generalized connections on closed manifolds $Q= \int Tr(F)$ in 2d, $Q= \int Tr(F \wedge F)$ in 4d, etc, is defined. 

- The space of generalized connections on a subdivided manifold can be calculated in terms of the spaces of generalized connections associated to its pieces. 
Thus, spaces of boundary connections can be computed from spaces associated to faces. 

- The soul of our generalized connections is a notion of higher homotopy parallel transport defined for smooth connections. We recover standard generalized connections by forgetting its higher levels. 


- The kth level of our higher gauge fields is trivial if and only if $\pi_{k-1} G$ is trivial. 
Then on dimension $2$ or higher $\bar{\cal A}^H_\Sigma \neq \bar{\cal A}_\Sigma$ when the gauge group is not simply connected. 
On the other hand, for $G=SL(2, {\mathbb C})$ or $G=SU(2)$ and $\dim \Sigma = 3$ we have $\bar{\cal A}^H_\Sigma = \bar{\cal A}_\Sigma$. 
Boundary data for loop quantum gravity is consistent with our space of generalized connections, but a path integral for quantum gravity with Lorentzian or euclidean signatures would be sensitive to homotopy data. 

\end{abstract}

\section{The space of generalized connections in quantum gauge theory and quantum gravity}
Ashtekar and Isham \cite{Ashtekar:1991kc} introduced the space of generalized connections modulo gauge transformations $\overline{\cal A / G}_\Sigma$ as the spectrum of the holonomy algebra, and it was proposed to be the quantum configuration space of gauge theories and gravity. 
For a more recent study of the spectrum of holonomy algebras see \cite{Abbati:2002mn}, which also considers algebras generated by non necessarily gauge invariant functions depending on the holonomies along finitely many loops, as is done in Baez's treatment of spin network states \cite{Baez:1994hx}.

General Relativity has a canonical frozen time formulation described in terms of Ashtekar-Barbero variables \cite{BarberoG:1994eia} where the fields are a $SU(2)$ connection (or gauge field) on the Cauchy surface $\Sigma$ and its conjugate momentum. This is now the basis of canonical loop quantum gravity. 

The space $\overline{\cal A / G}_\Sigma$ may be realized as a the space of equivalence classes of generalized connections (under generalized gauge transformations). 
A generalized connection $\bar{A} \in \overline{\cal A}_\Sigma$ 
is a groupoid homomorphism $A: P(\Sigma) \to G$ from a path groupoid associated to $\Sigma$ to $G$ (considered as a groupoid with one object), where the smoothness and even continuity conditions of a classical gauge field have been dismissed. Gauge transformations act by conjugation as usual. Thus, $\bar{A}$ is thought of as a parallel transport map.

\subsection{Kinematical canonical loop quantized gauge theory as a continuum limit}
Ashtekar and Lewandowski gave a ``projective construction'' of $\overline{\cal A / G}_\Sigma$ as the space of lattice gauge fields modulo gauge $({\cal A / G})_L= {\cal A}_L / {\cal G}_L$ for ``the finest lattice embedded in $\Sigma$. Since for any embedded lattice $L \subset \Sigma$ with $N_1$ edges the spaces ${\cal A}_L = G^{N_1}$ are measurable homogeneous spaces with a distinguished homogeneous measure, the projective construction made clear that the space $\overline{\cal A / G}_\Sigma$ may be constructed as a quotient of $\bar{\cal A}_\Sigma$, which is a measurable homogeneous space with a homogeneous measure, now called the Ashtekar-Lewandowski measure \cite{Ashtekar:1994mh}. 

The kinematical arena of canonical loop quantum gravity is ${\cal H}_\Sigma = L^2(\bar{\cal A}_\Sigma)$. In that space geometric operators corresponding to volumes of regions and areas of surfaces are defined, providing the notion of quantum geometry characteristic of loop quantum gravity. Also the constraints of canonical quantum gravity are defined on ${\cal H}_\Sigma$. 

A key property of ${\cal H}_\Sigma$ is that it hosts a representation of the diffeomrphism group of $\Sigma$. One could think that when the cutoff $L$ was present diffeomorphism symmetry was broken, but the continuum limit used to construct $\bar{\cal A}_\Sigma$ restores it. 
	
\begin{figure}
\centering
\includegraphics[width=0.3 \paperwidth]{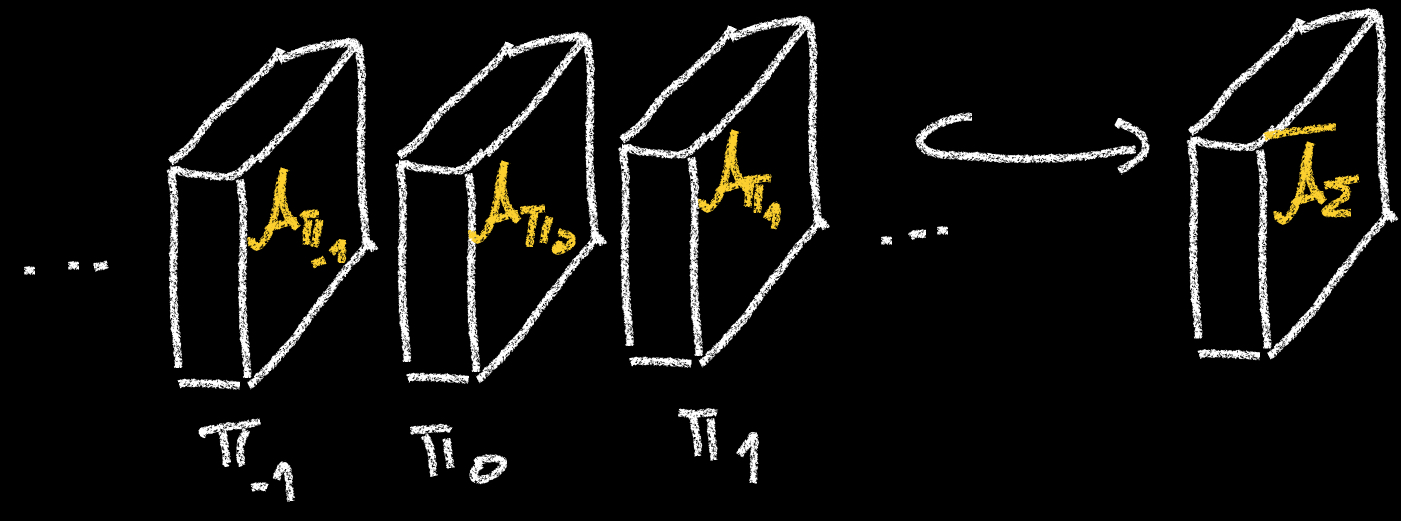}
    \caption{${\cal A}_\Sigma$ may have several connected components corresponding to different topological sectors, while $\bar{\cal A}_\Sigma$ is connected for connected gauge groups.}
    \label{TopologicalSectorsFIG}
\end{figure}

	The space of classical gauge fields 
 \[
 {\cal A}_\Sigma = \sqcup_\pi {\cal A}_\pi , 
 \]
 where the disjoint union is over the set of nonequivalent $G$ bundles over $\Sigma$, 
 is densely embedded in $\bar{\cal A}_\Sigma$. Thus, $\bar{\cal A}_\Sigma$ can be considered as a completion of ${\cal A}_\Sigma$. Notice, however, the disparity in sets of connected components shown in Figure \ref{TopologicalSectorsFIG}. 

\subsection{Covariant loop quantized theories as a continuum limit in the sense of renormalization
}
	On the other hand, in covariant loop quantum gravity most studies happen on $X$ a discretization of $M$ providing a cutoff for the path integral as it is customary in lattice gauge theory \cite{Creutz:1983njd}. An example is the EPRL model \cite{Engle:2007wy}. 

    A study of covariant loop quantum gravity in dimension 2 \cite{Oriti:2004qk} motivated the work presented in this article: On one hand, two dimensional gravity is known to be trivial because the Einstein Hilbert action is a topological invariant, and on the other hand, quantum two dimensional gravity can be constructed following the same strategy used to construct the EPRL model. The question was the result was a reasonable trivial quantum theory. The short answer is that the resulting quantum theory was not reasonable. The authors of \cite{Oriti:2004qk} also fixed the problem and obtained good model for two dimensional quantum gravity. The problem was not in the quantization strategy, the problem was that the gauge fields used in covariant loop quantization did not keep track of the topological information present in classical gauge fields. Certainly, the construction given in \cite{Oriti:2004qk} is related to the discretization of gauge fields described here, when restricted to the case of two dimensional bases and abelian gauge groups. 
 
	In an implementation of Wilsonian renormalization tailored to loop quantization \cite{Manrique:2005nn}, the continuum limit of lattice gauge theories is realized as the construction of a measure in $\bar{\cal A}_M$. This framework is consistent with the standard framework of lattice gauge theory, including the extensive work on computer simulations. 

Nowadays there are other proposals to implement coarse graining and renormalization within loop quantization. For a Hamiltonian approach see \cite{Lang:2017beo}; for the covariant (spin foam) approach see \cite{Asante2023} and references therein.

\section{Homotopy lattice gauge fields vs standard lattice gauge fields}
 The lattice cutoff induced by a lattice $L \subset \Sigma$ takes a smooth gauge field $A$ and stores only the information regarding the parallel transport maps along paths contained in $L$, producing $A^L$. 

 We introduced \cite{Orendain:2023tly, HLGF2} a finer cutoff based on a cellular decomposition $X$ of $\Sigma$. It takes $A$ and produces $A^{HL}$, a Homotopy Lattice Gauge Field (HLGF). $A^{HL}$ contains all the information in $A^L$ for the lattice $L=X_1$ (where $X_k$ denotes the k-skeleton of $X$), and it contains homotopical information that is not stored by $A^L$. 
 Consider a homotopy of curves $\gamma_t$ sharing source and target points (in $X_0$) interpolating between curves $\gamma_0, \gamma_1$ fitting in $X_1$. Additionally, demand that the homotopy fits in $X_2$. 
 The smooth gauge field $A$ produces a curve $A(\gamma_t) \in G$ starting at $A(\gamma_0)$ and finishing at $A(\gamma_1)$. The HLGF $A^{HL}$ does not store all the information in that curve, but it stores its homotopy class relative to fixed endpoints (which is the data contained in a standard lattice gauge field and the 1-dimensional information in $A^{HL}$).  This is the 2-dimensional information in $A^{HL}$. 
 
 A curve modulo thin homotopy equivalence is called a 1-globe, and a homotopy of curves, as described above, modulo thin homotopy equivalence is called a 2-globe. 

 The concept of thin homotopy equivalence is easy to implement in our context: Two 2-globes are considered equivalent if there is a homotopy (relative to their 0-dimensional source and target points) with image contained in $X_2$. In general, two k-globes are considered equivalent if there is a homotopy (rel. vertices) with image contained in $X_k$ relating them. 
 
 $A^{HL}$ associates points in $G$ to 1-globes in $X$, curves in $G$ up to homotopy relative to having fixed endpoints to 2-globes in $X$, and similar k-dimensional homotopies in $G$ are associated to (k+1)-globes in $X$. The mentioned data associated to k-globes needs to be consistent with the data associated to lower dimensional globes.

\begin{figure}
\centering
\includegraphics[width=0.35 \paperwidth]{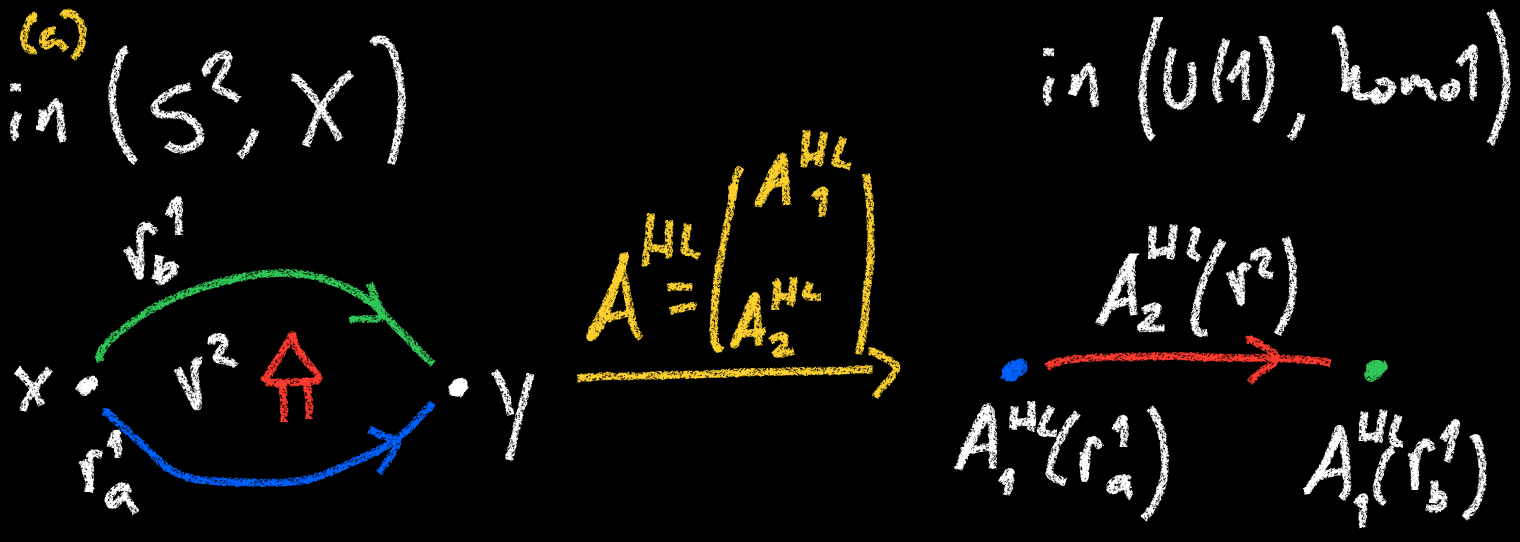}
\includegraphics[width=0.35 \paperwidth]{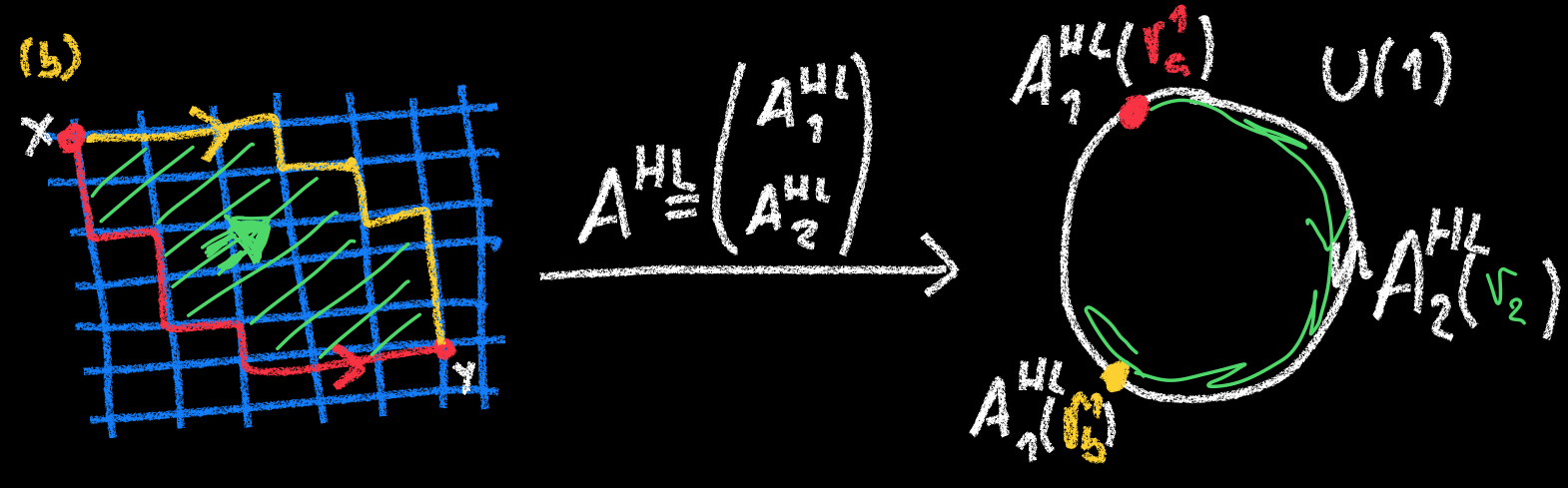}
    \caption{(a) In the left, a 2-globe in $X$ with its two 1-dimensional faces and its two 0-dimensional faces. 
In the right, the corresponding 1-globe in $G$. 
(b) Example for a 2-dimensional $X$ and $G=U(1)$. 
Notice that, in dimension 2 if $\pi_1 G$ were trivial $A_2^{HL}(\gamma^2)$ would carry no information. 
Also, in $\dim X = 2$ k-dimensional globes with $k>2$ are degenerate, which means that evaluation on higher dimensional globes carries only information contained in lower dimensional globes. }
    \label{HLGFs}
\end{figure}

 A standard lattice gauge field $A^L$ can be regarded as a groupoid homomorphism $A^L: P(L) \to G$ from the groupoid of paths fitting in $L \subset \Sigma$ to $G$ (considered as a groupoid with one object). A HLGF $A^{HL}$ is a homomorphism extending $A^L$ to higher dimensions that contain homotopical information%
 \footnote{The higher algebraic framework that we use was introduced in \cite{brown2007new} to solve the local to global problem in higher dimensional homotopy.}
\begin{equation}\label{HLGFeq}
    \rho^\bigcirc (X) 
\overset{A^{HL}}{\longrightarrow} \rho^\bigcirc (G)[-1] . 
\end{equation}
Left and right sides are towers of higher groupoids, where each level is consistent with the lower floors. The 1st floor is a groupoid; it has one operation for gluing, as appropriate for 1-dimensional things like paths. The 2nd floor is a 2-groupoid; it has two operations for gluing, as appropriate for a 2-dimensional tailing. 
But the structure needs to be such that the faces of the tiles and their gluing are described by the groupoid structure of the first level. 
Etc. In order to give precise algebraic rules for gluing in higher dimensions, one has to choose a ``prototypical shape''. One choice of shape is the cube; a k-dimensional cube has faces that are (k-1)-dimensional cubes. The choice that we are using in this work is called the globe.%
\footnote{In \cite{Orendain:2023tly} we used a formalism based on cubes, which contains the globular formalism as a substructure.}
1-globes are paths. k-globes interpolate between two (k-1)-globes as shown in the left of Figure \ref{HLGFs}. $A^{HL}$ acts on k-globes in $X$ and produces (k-1)-globes in $G$. The mismatch in  the dimension of the globes received and the globe produced (indicated by the $[-1]$ in \eqref{HLGFeq}) is not a problem, what is important for a homomorphism is to match the gluing operations and since $G$ is not only a manifold, but it has a product, this operation is matched with the concatenation of globes that share a 0-dimensional face. We are familiar with this in the case of 1-globes (i.e. paths); recall that the parallel transport along a path $\gamma$ is encoded by the group element $A(\gamma)$ as shown in Figure \ref{HLGFs}. 
Because we are considering gauge fields with a cutoff, $A^{HL}$ acts on globes in $X$ considered up to thin homotopy (relative to the vertices)
and produces globes in $G$ up to homotopy ({\em not thin}) relative to vertices corresponding to the data associated to 1-dimensional globes. 
Because of this equivalence of globes in the group up to homotopy, it is common that at higher levels the information in $A^{HL}$ is trivial. 
At the level of k-globes in $X$ the information in $A^{HL}$ is always trivial if and only if $\pi_{k-1} G$ is trivial. Then for base spaces of dimension $d \geq 2$ the HLGF $A^{HL}$ contains more information than the standard lattice gauge field $A^L$ whenever $G$ is not simply connected. On the other hand, for $G=SU(2)$ and a base of dimension $d \leq 3$ the information in the field $A^{HL}$ is all contained in the first level (which we may call $A^L$). In the cases of $SU(2)$ and $SL(2, {\mathbb C})$ the nontrivial information in higher levels of $A^{HL}$
starts to show up at dimension $d=4$. 

Gauge transformations act on HLGFs by conjugation: 
Consider a smooth gauge field $A$ acting on a (k-1)-dimensional homotopy of curves $\gamma_t$ sharing endpoints. A gauge transformation acts on $A(\gamma_t)$ by conjugation transforming it into 
$A'(\gamma_t)= (g(y))^{-1} A(\gamma_t) g(x)$, where $x = s(\gamma_t)$ and $y = t(\gamma_t)$ for all $t$. 
The independence of the gauge transformation on the homotopy parameter is a big advantage of working with globes. The homotopy lattice cutoff takes a smooth gauge field $A$ and transforms it into $A^{HL}$; since the cutoff is based on globes (fitting in $X$), a gauge transformation acts on $A^{HL}(\gamma)$ by simple conjugation.

\begin{figure}
\centering
\includegraphics[width=0.5 \paperwidth]{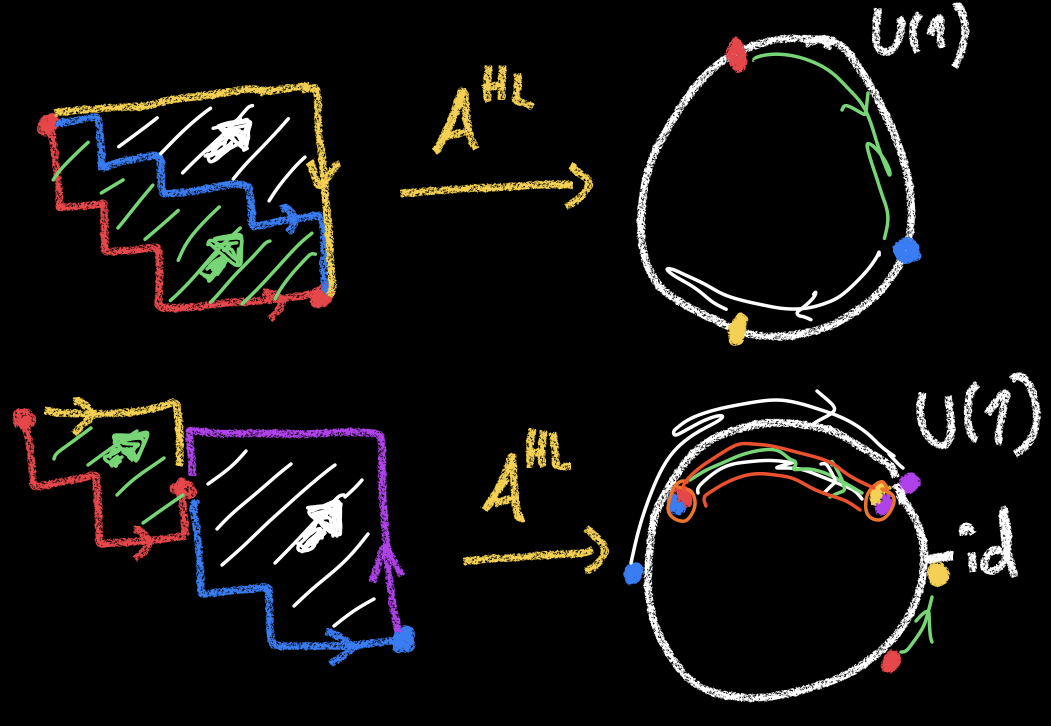}
    \caption{First, two 2-globes are glued along a 1-globe that they share. In $G$, $A^{HL}$ yields two curves (up to homotopy rel. endpoints) that can be concatenated. Second, two 2-globes are glued along a 0-globe (vertex) that they share; these 2-globes correspond to two homotopies of curves $\gamma_t , \eta_t$ in $X$ that may be composed to produce a new homotpy of curves $\delta_t = \eta_t \gamma_t$. Evaluation of a gauge field on a concatenation of curves corresponds to the product of the evaluations. 
    In $G$, $A^{HL}$ yields two (classes of) curves. The pointwise product of those curves yields a curve; the class of that curve does not depend on the representatives.} 
    \label{Gluing}
\end{figure}

A standard lattice gauge field $A^L$ is stored in the computer by evaluating it on the set of lattice links. Parallel transport along a general path in the lattice can be calculated from the group elements associated to the links contained in the path. Similarly, a HLGF $A^{HL}$ can be stored in the computer by evaluating it on a set of generators. For example, the set of generators needed to construct any 2-globe can be chosen to be one 2-globe for each lattice plaquette together with the set of degenerate 2-globes with image contained in lattice links. 
Parallel transport along general k-globes in $X$ can be calculated from the data associated to generators; see Figure \ref{Gluing}. 
Since k-globes can be glued using $k$ different gluing operations, it is crucial that gluing following different orders yields the same answer. This property of higher groupoids is called the interchange rule. This rule is part of the structure of 
$\rho^\bigcirc (X)$ and $\rho^\bigcirc (G)[-1]$, and since $A^{HL}$ is a morphism we can store $A^{HL}$ in a computer evaluating it in a set of generators. 

A complication is that there may be relations among the generators; this means that the set of evaluations needs to satisfy compatibility conditions. 
Consider for example the 3-torus $\Sigma = T^3$ as base space with a cellular decomposition by a cartesian cubiculation $X$ with gauge group $G=U(1)$. The generating set of 1-globes can be chosen to be the set of links; that set of generators is an independent set. Now add one generator for 2-globes for each plaquette; a particular 2-globe needs to be chosen, but the rest of the argument does not require us to choose one. The set of generators that includes the edges and the plaquettes is not an independent set. We can freely determine first the evaluations of 1-dimensional generators; then the evaluation on each plaquette carries only the discrete information of a homotopy class of curves relative to fixed endpoints in $U(1)$. Are the six evaluations on the plaquettes binding a 3-cube independent? No. 
The six generators of 2-globes in $X$ are independent, but evaluation of $A^{HL}$ takes them to curves in $U(1)$ up to homotopy (rel. to fixed endpoints) where they are not independent. Here is one explanation of their dependence: There is a relationship between the generators of 2-globes associated the boundary feces of a 3-cube and the generator of 3-globes associated to the 3-cube itself. That generator of 3-globes is mapped by $A^{HL}$ to a 2-globe in $U(1)$, but all 2-globes in $U(1)$ are degenerate because $\pi_2 U(1)$ is trivial. Then there is a relationship between the $A^{HL}$ evaluations of the generators of 2-globes associated the boundary feces of a 3-cube. 
In each particular situation an independent choice of generators may be chosen. For more details see \cite{HLGF2}. 

The description of HLGFs in terms of evaluation in a set of generators (without the algebraic framework needed for gluing) was proposed earlier \cite{Meneses:2017vqn, Meneses:2019bok} with motivations in differential topology and physics. 
In that work, it was proven that, similarly to what happens in the continuum \cite{Barrett:1991aj}, a gauge field $A^{HL}$ determines a $G$-bundle over the base space $(\Sigma, X)$, and a connection in the corresponding bundle (up to homotopy relative to the information contained in the lattice gauge field). 

The formulation of HLGFs in terms of the higher algebraic language designed to talk about higher homotopy solves a crucial physics problem: Coarse graining in this framework is done by gluing, and the interchange rule of higher groupoids tell us that gluing in different orders leads to the same result. 

\begin{figure}
\centering
\includegraphics[width=0.5 \paperwidth]{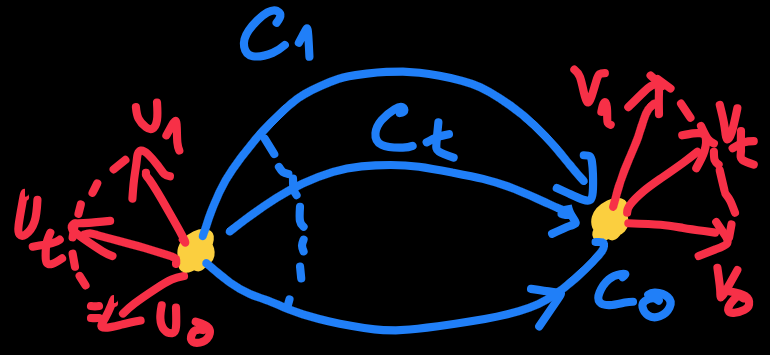}
    \caption{The higher parallel transport along a 2-globe: A homotopy of initial conditions on the 0-dimensional source of the globe is parallel transported by a homotopy of curves (a representative of the 2-globe). The result is a homotopy of final conditions at the 0-dimensional target of the globe which, up to homotopy relative to endpoints, is independent of the choices of representatives.} 
    \label{HigherParTranspFig}
\end{figure}

Another property that is manifest because of using the tools of non abelian algebraic topology is a picture of higher parallel transport. Higher homotopy parallel transport is a feature of smooth gauge fields. Consider a homotpy of curves $\gamma_t$ in the base sharing source and target points $x= s(\gamma_t)$, $y= t(\gamma_t)$. A smooth gauge field $A$ acting on the homotopy of curves yields a curve $A(\gamma_t)$ in $G$ starting at $A(\gamma_0)$ and finishing at $A(\gamma_1)$. This information can be used to parallel transport a homotopy of initial conditions $u_t$ in the fiber over $x$; the result is
$v_t = A(\gamma_t) u_t$ a corresponding homotopy of final conditions in the fiber over $y$ (see Figure \ref{HigherParTranspFig}). 

The HLGF $A^{HL}$ induced by $A$ cannot parallel transport the entire curve $u_t$ in the fiber over $x$, but it can transport the class of the curve considered up to homotopy relative to having the endpoints fixed. The mentioned parallel transport of the endpoints of the homotopy of initial conditions is fixed by the data on the first level of $A^{HL}$, the data on the standard lattice field. 
Analogously, $A^{HL}$ acting on a k-globe (a(k-1)-dimensional homotopy of curves fitting in $X$ sharing endpoints considered up to thin homotopy) $[u]$ can parallel transport initial data over he 0-dimensional source $x$ to the 0-dimensional target $y$, and the appropriate data being transported is a (k-1)-dimensional globe in $G$ considered up to homotopy (not thin homotopy) relative to having its 0-dimensional source and target points fixed. 
\[
[v] = A^{HL}(\gamma) [u] . 
\]
For further details on this homotopy higher parallel transport see \cite{Orendain:2023tly, HLGF2}. 

An important property of an HLGF $A^{HL}$ is that it determines a $G$-bundle over $\Sigma$ up to bundle equivalence. In fact, this was one of the most important motivations for searching for an appropriate extension of the standard notion of gauge field on the lattice. 
A $G$-gauge field $A$ in the continuum on a base space $\Sigma$ determines a bundle. A sketch of the  construction in \cite{Barrett:1991aj} is as follows: 
In an atlas for $\Sigma$, construct a set of local trivializations using $A$ (and a radial gauge condition on each chart); then the transition functions become functions of $A$, determining a $G$-bundle over $\Sigma$. 
After the standard lattice cutoff, the lattice gauge field $A^L(A)$ determines the evaluation of the mentioned transition functions only on a discrete set of points, which is not enough to determine a bundle class. 
The notion of extended lattice gauge fields proposed in \cite{Meneses:2017vqn, Meneses:2019bok} arises from a refined lattice cutoff of a gauge field $A$ in the continuum where, a part from the data on $A^L(A)$, the extended lattice gauge field keeps the missing information regarding the mentioned transition functions determined by $A$, but only up to homotopy relative to vertices (which correspond to the data determined by $A^L(A)$). The resulting extension of a lattice gauge field associates data to cells of all dimensions in a cellular decomposition, not just to cells of dimension 1. 
Then it is verified that the resulting extended lattice gauge field does determine a bundle up to bundle equivalence. 
This result is almost tautological within the construction given in 
\cite{Meneses:2017vqn, Meneses:2019bok}, and it applies to HLGFs due to the results of \cite{Orendain:2023tly} showing that a HLGF evaluated on an appropriately chosen set of generators yields an extended lattice gauge field as defined in \cite{Meneses:2017vqn, Meneses:2019bok}.

\section{The space of HLGFs}
	The space of HLGFs of $X$ is a covering space of the space of standard LGFs%
\begin{equation} \label{CoveringMapX}
    {\cal A}^{HL}_X \to {\cal A}^L_{L=X_1} \simeq G^{N_1} . 
\end{equation}
If the higher levels of the HLGF are trivial \eqref{CoveringMapX} is a diffeomorphism. 
In the general case the covering map has a discrete fiber on which a finite group acts freely transitively. This group is constructed explicitly in terms of generators and relations \cite{Meneses:2017vqn}. 

\subsection{Measures in ${\cal A}^{HL}_X$}

Since the space of standard lattice gauge fields is a homogeneous geometric space, 
the covering map \ref{CoveringMapX} tells us that ${\cal A}^{HL}_X$ is a locally homogeneous geometric space. Then ${\cal A}^{HL}_X$ is a measurable space. 

Recall that gauge transformations act on HLGFs by simple conjugation by group elements associated to the discrete set of vertices $X_0$. 
Then the space of gauge orbits $({\cal A / G })^{HL}_X$ is simply calculated from ${\cal A}^{HL}_X$ as 
$({\cal A / G })^{HL}_X = {\cal A}^{HL}_X/{\cal G}_{X_0}$. The map sending each HLGF to its class 
${\cal A}^{HL}_X \to {\cal A}^{HL}_X/{\cal G}_{X_0}$ tells us that $({\cal A / G })^{HL}_X$ is also a measurable space and measures in ${\cal A}^{HL}_X$ are push forward to $({\cal A / G })^{HL}_X$. 

There is a Haar measure in ${\cal A}^{HL}_X$ induced locally by the Haar measure in ${\cal A}^L_{L=X_1}$%
\footnote{Notice, however, that if the covering map \eqref{CoveringMapX} has fibers with infinite cardinality, the resulting Haar measure would not be normalizable.}. 
Then physical measures can be constructed multiplying by appropriate Boltzmann factors. 

Heat kernel measures for standard lattice gauge fields are constructed 
calculating the Boltzmann factor of a given lattice gauge field $A^L$ 
as a product over plaquettes; the factor associated to a $A$ and a plaquette $\Box$ is computed 
using solutions of the heat equation 
in $G_\Box$, a copy of $G$ associated to the plaquette $\Box$. 
Let us loosely call $\partial\Box$ a loop binding the plaquette, 
and let $K(t,g)$ denote a solution of the heat equation in $G$ with distributional source at $id \in G$ after time $t$. Then the Boltzmann factor 
in the notation used in \cite{Creutz:1983njd} 
is $e^{S_\Box(\beta, A^L)} = K(1/\beta, A^L(\partial \Box))$. 

For HLGFs, we define heat kernel measures following the same procedure, but with one single difference: the Boltzmann factor associated to each plaquette is calculated solving the heat equation in $\tilde{G}$ the universal covering group of $G$. 
Let $\tilde{K}(t, \tilde{g})$ denote a solution of the heat equation in $\tilde{G}$ with distributional source at $id \in \tilde{G}$. 
Also, let us loosely write $\Box$ for a 2-globe with image in $\Box$ and corresponding to the class of a homotopy of loops growing from the constant loop at one vertex of the plaquette to a loop that goes around it once.  
A HLGF $A^{HL}$ acting on $\Box$ yields a curve in $G$ starting at $id \in G$ and ending at the holonomy along the plaquette’s boundary. When this curve is considered up to homotopy relative to fixed endpoints we obtain an element of $\tilde{G}$ that we loosely denote by $A^{HL}(\Box)$. 
Then the Boltzmann factor for HLGFs 
is $e^{S_\Box(\beta, A^{HL})} = \tilde{K}(1/\beta, A^{HL}(\Box))$. 

Notice that if $G$ is simply connected, the expression written for the Boltzmann factor is identical to the expression for the standard lattice gauge field $A^L(A^{HL})$. 

Since $K(t,g)$ is invariant under the adjoint action of $G$ on itself, the proposed heat kernel measure is gauge invariant. 

BF measures are measures with support concentrated in the moduli space of flat connections. 
A concrete implementation, employed for standard lattice gauge fields,  is to use delta functions on $G$ as Boltzmann factors to demand that the parallel transport on contractible loops be trivial. 
Our implementation for HLGFs  complements that requirement with local conditions on homotopy data located at higher levels of the HLGF. Now we describe these conditions: 
Recall that we can describe the HLGF by evaluation on a set of generators, and that each of the generators of k-globes is associated to one k-dimensional cell. We can select a set of generators with the property that (for $k>1$) each k-globe has a single vertex as its 0-dimensional source and 0-dimensional target. Our 2-globe generators are then bounded by a loop. The implementation of a  BF measure of standard lattice fields requires that the parallel transport along the binding loop be the identity. Since our generator of 2-globes represents a homotopy between the constant loop based at the base vertex and the generator, a flat connection would map that homotopy to the constant curve in $G$ equal to the identity. Thus, for a choice of generators satisfying the condition described above, the condition on the evaluation on k-globes $\gamma$ (for $k>1$) is that $A^{HL}(\gamma)$ be a degenerate (k-1)-globe on $G$ with image the identity in $G$. 

The BF measure in ${\cal A}^{HL}_X$ 
just described has the following properties: (i) It is gauge invariant. (ii) Its support is entirely contained in the topological sector admitting flat connections.

\subsection{Relation between ${\cal A}^{HL}_X$ and the space of smooth connections}

The results of \cite{Meneses:2017vqn, HLGF2} described at the end of the previous section imply that the homotopy lattice cutoff is a continuous map 
\[
{\cal A}^\infty_\Sigma = \sqcup_\pi {\cal A}^\infty_\pi \to {\cal A}^{HL}_X
\]
sending $A$ to $A^{HL}(A)$, where both $A$ and $A^{HL}(A)$ determine the same bundle up to equivalence.

\section{Coarse graining and continuum limit}
Given two cellular decompositions $X \leq X'$ of $\Sigma$, where $X'$ is finer than $X$, a coarse graining map 
\begin{equation} \label{CoarseGrMap}
    {\cal A}^{HL}_{X'} \to {\cal A}^{HL}_X 
\end{equation}
follows directly from the algebraic expressions for gluing globes from $X'$ to make globes in $X$, and the fact that $A^{HL}_{X'}$ is a homomorphism \cite{Orendain:2023tly}. 

The above coarse graining map, specialized to the 1-dimensional data, tells us that a standard lattice gauge field of a finer lattice induces a standard lattice gauge field on a coarser lattice. 
Using these coarse graining maps, the space of generalized connections of loop quantized theories 
is constructed as a limit of spaces of gauge fields on the lattice in which the embedded lattice $L\subset \Sigma$ becomes as fine as possible \cite{Ashtekar:1994mh}. This space should be considered the result of a kinematical removal of the cutoff. 
The same procedure can be used to construct $\bar{\cal A}^H_\Sigma$, the space of homotopy aware generalized connections 
\begin{equation}\label{ProjLim}
    \bar{\cal A}^H_\Sigma \doteq \underleftarrow{\lim}_{X \to \Sigma}  \ {\cal A}^{HL}_X . 
\end{equation}
A generalized homotopy aware connection 
$A \in \bar{\cal A}^H_\Sigma$ induces 
$A^{HL}(A) \in {\cal A}_X^{HL}$ for any $X$. 

For definition \eqref{ProjLim} to make sense, it is necessary that the set cellular decompositions  considered be partially ordered and directed. This property is easily verified in the piecewise linear category. 

\subsection{Measures in $\bar{\cal A}^H_\Sigma$}

The construction described above implies that $\bar{\cal A}^H_\Sigma$ is a locally homogeneous measurable space. An assignment of a measure to each of the spaces ${\cal A}^{HL}_X$ that is compatible with coarse graining defines a measure in $\bar{\cal A}^H_\Sigma$. 

A homogeneous measure on $\bar{\cal A}^H_\Sigma$ may not exist, even for compact gauge groups. 
As mentioned earlier, if the covering map \eqref{CoveringMapX} has fibers with infinite cardinality, the Haar measure on the spaces ${\cal A}^{HL}_X$ is not normalizable. This would imply that when calculating the expectation value of any cylindrical function $f$, integrating over all the degrees of freedom on which $f$ does not depend, the result would be infinite. 
This happens whenever 
$|\pi_k G|= \infty$ for some $k < \dim \Sigma$. 

In general, physical measures would not be calculable as a product of a homogeneous measure and a Boltzmann factor. 
This should not be seen as an obstacle for the utility of $\bar{\cal A}^H_\Sigma$ in physics because physical measures in the continuum are supposed to be constructed as a continuum limit following Wilsonian renormalization. A physical measure in $\bar{\cal A}^H_\Sigma$ is determined by a collection of completely renormalized measures at the spaces ${\cal A}^{HL}_X$; by construction, a collection of completely renormalized measures is compatible with coarse graining maps. 

It is easy to verify that a collection of BF measures, as described in the previous section for different cellular decompositions $X$, is compatible with coarse graining maps. That is, the measures are already completely renormalized and they define a measure in $\bar{\cal A}^H_\Sigma$ by a projective limit. 

As expected, the construction of a measure for Yang-Mills theory (in $d>2$)using the heat kernel measures defined in the previous section requires Wilsonian renormalization. 
In the case of two-dimensional Yang Mills theory the measures described in the previous section are compatible with projection leading to a measure in $\bar{\cal A}^H_\Sigma$. For details see \cite{HLGF2}.

\subsection{$\bar{\cal A}^H_\Sigma$ is a completion of ${\cal A}^\infty_\Sigma$ preserving topological sectors}

The space of smooth connections 
\[
{\cal A}^\infty_\Sigma = \sqcup_\pi {\cal A}^\infty_\pi 
\quad \mbox{ is densely embedded in } \quad 
\bar{\cal A}^H_\Sigma = \sqcup_{cc} \bar{\cal A}^H_{cc},
\]
where $cc$ labels the connected component, 
and two fields $A, A' \in \bar{\cal A}^H_\Sigma$ in the same $cc$ determine equivalent bundles. 

The dense embedding factorizes in topological sectors 
${\cal A}^\infty_\pi \hookrightarrow \bar{\cal A}^H_{cc(\pi)}$. Because of this property, the space $\bar{\cal A}^H_\Sigma$ may be considered a completion of ${\cal A}^\infty_\Sigma$, and since the embedding preserves topological sectors $\bar{\cal A}^H_\Sigma$ is a refinement of the usual space of generalized connections $\bar{\cal A}_\Sigma$.

\subsection{Relation between $\bar{\cal A}^H_\Sigma$ and $\bar{\cal A}_\Sigma$}

The collection of maps \eqref{CoveringMapX} for every $X$ cell. dec. of $\Sigma$ yields a covering map 
\[
\bar{\cal A}^H_\Sigma \to \bar{\cal A}_\Sigma . 
\]

As mentioned earlier, $\bar{\cal A}^H_\Sigma = \bar{\cal A}_\Sigma$
if $\pi_k G$ is trivial for all $k < \dim \Sigma$. 
In the case of gravity, relying on the gauge group $G=SL(2, {\mathbb C})$ or $G=SU(2)$ we would have that at $\Sigma$ a three-dimensional Cauchy surface $\bar{\cal A}^H_\Sigma = \bar{\cal A}_\Sigma$. 

As soon as we couple a Maxwell field, however, the space of homotopy aware generalized connections $\bar{\cal A}^H_\Sigma$ carries more information; that extra information would have a definite impact on the state space and on 
the dynamics, and it may be essential to appropriately couple fermions. 

Also in the case of gravity, but (covariantly) considering dynamics HLGFs are not equivalent to standard lattice gauge fields because the propagator computed at four dimensions would be sensitive to the extra information on HLGFs. 

If more than one topological sector is physically relevant, as is the case in Yang-Mills theory, a path integral on $\bar{\cal A}^H_M$ 
(computed as a continuum limit in the sense of renormalization) would consider each sector. 

If only one sector is supposed to contribute to the path integral, as is the case in gravity where the frame bundle is the only $G$-bundle relevant, that truncation of the space $\bar{\cal A}^H_M$ is easily done, and it can be done at a local level. 
As previously mentioned, in \cite{Oriti:2004qk} a model for two-dimensional quantum gravity was constructed following a {\em corrected} spin foam quantization related to HLGFs.

\section{A simple formula for the topological charge}

As mentioned earlier, a HLGF determines a $G$-bundle over the base manifold. Thus, we know that there is a continuous map assigning a topological charge to each HLGF $A^{HL}$. In this section we will see that the algebraic tools of HLGFs yield a is simple formula for that map. 

The topological charge is a function of the gauge field usually presented as an integral of a polynomial in the curvature. 
In the cases of the topological charge of a $U(1)$ connection on $S^2$ and to the topological charge of a $SU(2)$ connection on $S^4$, 
a differential geometric result (see for example \cite{dubrovin1984modern}) is that the topological charge can also be computed from the winding number of the map from the sphere to the sphere. The construction of the map is as follows: Consider an atlas of the sphere with two charts (one over the Southern Hemisphere and the other over the Northern Hemisphere), with an overlapping region that is a neighborhood of the Equator. The map is the transition function restricted to the Equator $\phi_{SN}|_{Eq}$. 

The characterization of the topological charge as a winding number is readily calculable in terms of HLGFs. Moreover, the spirit of the calculation is highly analogous to the construction of extended lattice gauge fields in \cite{Meneses:2019bok}. In that reference the topological charge is calculated as a physical observable computed in terms of ``relational observables'', and in that reference we had to assume that the base was covered by only two simplices of highest dimension. 

\begin{figure}
\centering
\includegraphics[width=0.5 \paperwidth]{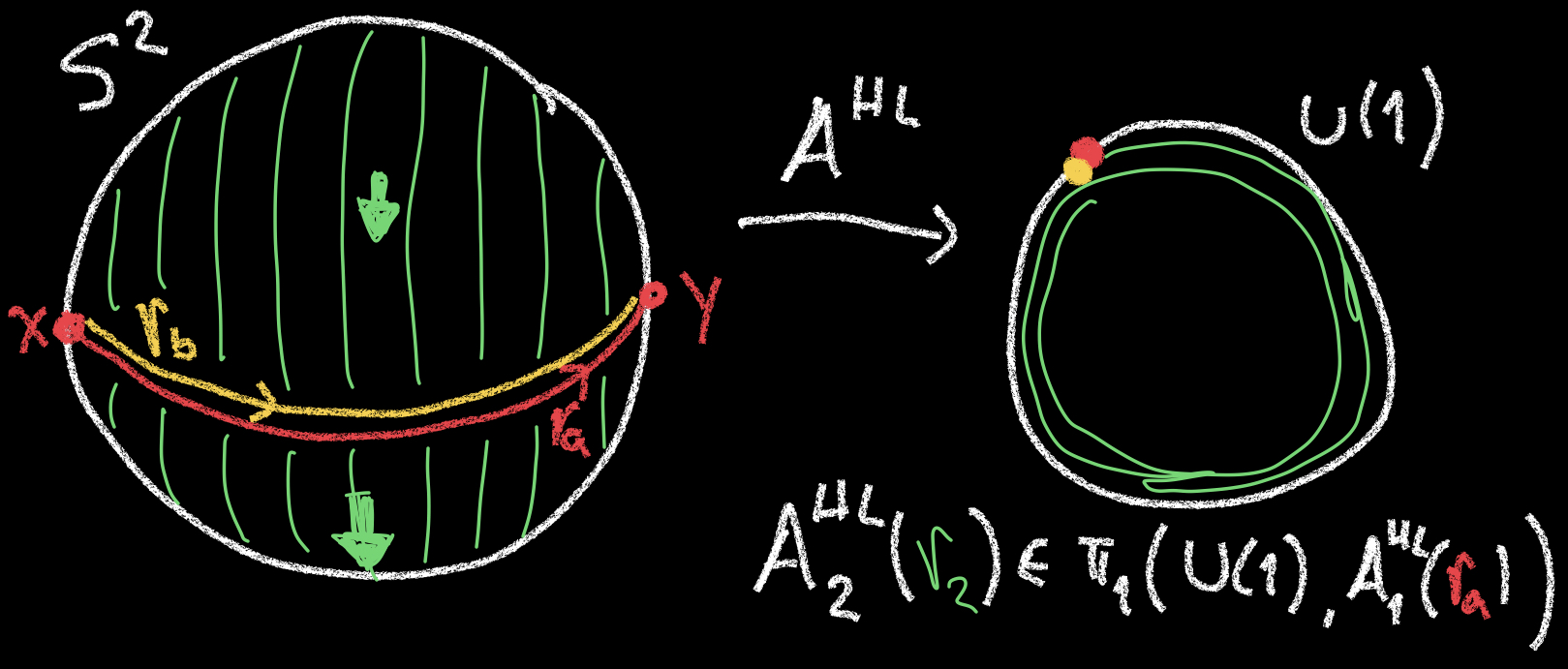}
    \caption{In $S^2$, the shown 2-globe corresponds to a homotopy of curves from $x$ to $y$ starting at the red curve, finishing at the yellow curve and winding around $S^2$ once. A gauge field $A$ transforms that homotopy of curves in the base into a curve in $U(1)$ starting at the parallel transport associated to the red curve and finishing at the parallel transport associated to the yellow curve. Since the red and the yellow curve coincide, the HLGF $A^{HL}(A)$ evaluated on the 2-globe yields a winding number.} 
    \label{TopChargeAsWinding}
\end{figure}

In terms of HLGFs, the gluing properties of globes let us glue the globes associated to all the cells covering a hemisphere into a single globe of highest dimension. Then the formula for the topological charge as a winding number of the transition function between two charts applies. Moreover, that formula can be simplified by gluing the two globes corresponding to the two hemispheres into a single globe. Then the winding number can be pictured as explained in Figure \ref{TopChargeAsWinding}, and 
the resulting formula for the topological charge is simply 
\[
Q(A^{HL})= A_2^{HL}(\gamma_d) \ \ \mbox{if} \ \  [\Phi(\gamma_d)]_{\pi_d} = 1 . 
\]

\section{Local to global}

	Using globes in a space with a cellular decomposition one can capture all the information present in the homotopy groups of that space. That is the reason for calling the higher algebra of gluing globes the fundamental infinity groupoid of that space. Its algebraic properties lead to the proof of the Higher Homotopy Seifert-van Kampen Theorem solving the local to global problem in homotopy: When a space with a cellular decomposition is chopped into pieces that are properly subdivided by the decomposition, the groupoid associated to the whole space can be calculated in terms of the groupoids associated to the pieces. This result was the original goal of Brown and collaborators for developing their framework to study higher homotopy by means of higher category theory \cite{brown2007new}. 

\begin{figure}
\centering
\includegraphics[width=0.5 \paperwidth]{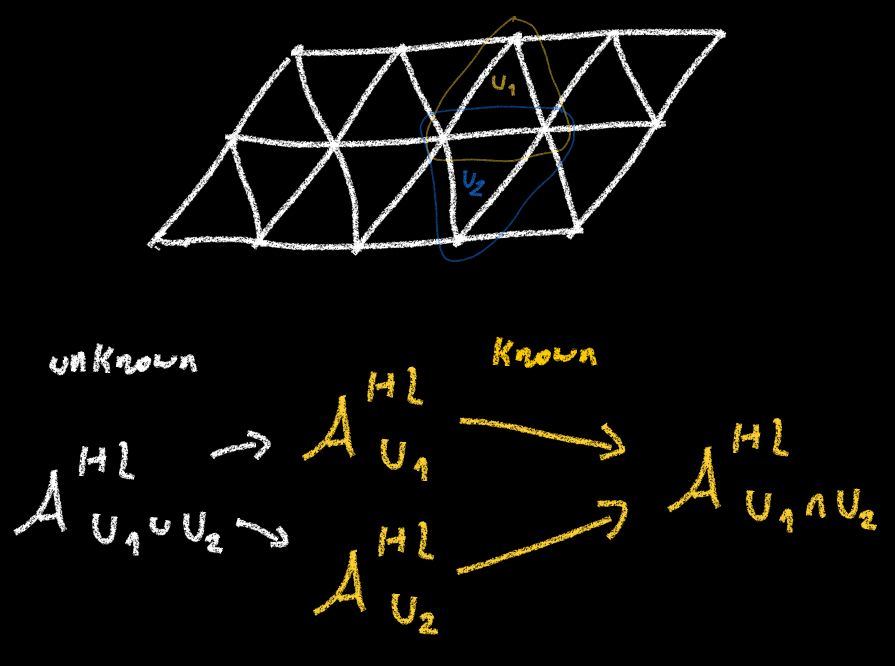}
    \caption{The Higher Homotopy Seifert-van Kampen Theorem applied to $\bar{\cal A}^H_\Sigma$. We can use as data the space of HLGFs restricted to the sets of a cover and recover the space of HLGFs in the whole space from them, provided that compatibility conditions are satisfied at intersections. } 
    \label{LocalToGlobal}
\end{figure}

	That result applied to HLGFs implies that given discretization (cellular decomposition) and an open cover of the space compatible with the decomposition, we can calculate the space of gauge fields associated to the whole discretized space in terms of the spaces of fields associated to the discretized pieces. See Figure \ref{LocalToGlobal}

	This local factorization survives the kinematical continuum limit used to build $\bar{\cal A}^H_\Sigma$. For details see \cite{HLGF2}. 

	If we use $\bar{\cal A}^H_\Sigma$ to construct a kinematical space of states associated to $\Sigma$, then we can construct that Hilbert space in terms of the spaces of states $\bar{\cal A}^H_U$ associated to the pieces. This is relevant in particular when the hypersurface is the boundary of a spacetime region and that boundary comes subdivided into faces. Importantly, the mentioned factorization property works for HLGFs and not for HLGFs modulo gauge: one has to glue the pieces first and later consider gauge orbits if one wants.

\section*{Acknowledgments}
Supported by grant PAPITT-UNAM IN114723 

\medskip
 
\bibliographystyle{unsrt}

\bibliography{references}

\end{document}